\documentclass[11pt]{article}             
\usepackage{osid}                         %
\usepackage{color,graphicx,bbold}
\usepackage{amsmath,amssymb,amsthm}
\usepackage{epsfig}

\newcommand\bra[1]{\left\langle#1\right|}
\newcommand\ket[1]{\left|#1\right\rangle}
\newcommand\be{\begin{equation}}
\newcommand\ee{\end{equation}}
\newcommand{\modulo}[1]{\left\vert#1\right\vert}
\newcommand\braket[2]{\left.\left\langle#1\right|#2\right\rangle}
\newcommand{\vett}[1]{\mbox{\boldmath $#1$}}
\newcommand{\en}{\varepsilon_{n}}

\title{Decoherence in a fermion environment: Non-Markovianity and Orthogonality Catastrophe}
\author{F. Plastina \\{\footnotesize\it Dipartimento di Fisica, Universit\`a della Calabria, 87036 Arcavacata di Rende (CS), Italy\\
INFN sezione LNF-Gruppo collegato di Cosenza, Italy \& francesco.plastina@fis.unical.it}\\[2ex]
        A. Sindona   \\{\footnotesize\it Dipartimento di Fisica, Universit\`a della Calabria, 87036 Arcavacata di Rende (CS), Italy\\
INFN sezione LNF-Gruppo collegato di Cosenza, Italy } \\ [2ex]
J.~Goold
\\{\footnotesize\it Clarendon Laboratory, University of Oxford, United Kingdom\\
Physics Department, University College Cork, Cork, Ireland} \\
[2ex] N.~Lo~Gullo \\{\footnotesize\it Dipartimento di Fisica e
Astronomia Galileo Galilei and CNISM, Universit\`{a} di Padova,
Via Marzolo 8, 35122 Padova, Italy} \\ [2ex] S.
Lorenzo \\{\footnotesize\it Dipartimento di Fisica, Universit\`a della Calabria, 87036 Arcavacata di Rende (CS), Italy\\
INFN sezione LNF-Gruppo collegato di Cosenza, Italy } }

\begin{document}

\maketitle
\begin{abstract}
We analyze the non-Markovian character of the dynamics of an open
two-level atom interacting with a gas of ultra-cold fermions. In
particular, we discuss the connection between the phenomena of
orthogonality catastrophe and Fermi edge singularity occurring in
such a kind of environment and the memory-keeping effects which
are displayed in the time evolution of the open system.
\end{abstract}

\section{Introduction and Motivations}
The precise definition and quantitative description of memory
effects (or non-Markovianity) have become a central issue in the
theory of open quantum systems
\cite{uno,blp,Chruscinski2010,rhp,sun,erica,vasile,luo,due,spettro,rev},
and have been the subject of recent experimental efforts
\cite{experiments}. It has been argued that non-Markovianity has
the potential of being exploited to pursue new quantum
technologies \cite{due},  and that it can be thought as a resource
in quantum metrology \cite{tre}, for the generation of entangled
states \cite{entgeneration} and for quantum key distribution
\cite{quattro}.

Various characterizations of the non-Markovian behavior have been
given, which capture different aspects of the decohering dynamics
of an open system. They include the lack of divisibility of the
map describing the time evolution of the system of interest
\cite{uno,rhp}, or the back-flow of the information that the
system itself had previously lost, described either in terms of
the distinguishability of the evolved states \cite{blp} or of the
quantum Fisher information \cite{sun}. Further proposals have been
put forward, based on the decay rates entering the master equation
\cite{erica}, on the use of the quantum mutual information
\cite{luo} or of channel capacities \cite{due}, on spectral
considerations \cite{spettro}, and on the temporary expansion of
the volume of the states accessible through the reduced dynamics
\cite{det}. Moreover, many other properties, related to the
locality \cite{sei}, to the complexity \cite{znidaric} or to the
size of the environment \cite{lorenzoPRA}, have been investigated
in different physical settings, ranging from spin systems
\cite{apollaro11,haikkagold} to Bose-Einstein condensates of
ultra-cold atoms \cite{haikkabec}.

Despite the conceptual differences,  several quantifiers of the
amount of non-Markovianity give similar qualitative (and sometimes
also quantitative) descriptions when applied to the dynamics of
simple quantum systems such as a qubit\cite{altricin,haikkagold}.
In particular, this holds true for the specific case of a purely
dephasing dynamics where the system loses coherence due to its
interaction with the environment, without any energy exchange. In
this case, indeed, the open system evolution is completely
characterized by a so called decoherence factor, which is the only
ingredient necessary to evaluate the amount of non-Markovianity.

Specifically, let us consider a two-level system (a qubit, with
energy eigenstates $\ket{\alpha}$, $\alpha=0,1$) interacting with
its environment in such a way as to preserve its energy. This
implies that the interaction Hamiltonian commutes with that of the
qubit and, as a result, the qubit state can be written as \be
\rho(t) = \begin{pmatrix}\rho_{00}(0) & \nu(t) \rho_{01} (0) \\
\nu^*(t) \, \rho_{10}(0) & \rho_{11}(0) \end{pmatrix}, \quad
\nu(t)=Tr_{env} \{e^{i H_{1} t} \, e^{-i H_{0} t} \, \rho_{env}\}
\, ,\label{tev} \ee where $\rho_{env}$ is the initial state of the
environment, while $H_{\alpha} = H_{env} + \bra{\alpha} H_{int}
\ket{\alpha}$ are effective Hamiltonians for the environment,
conditioned on the state of the qubit ($\alpha =0,1$).

If the initial state of the environment is pure, $\rho_{env} =
\ket{\phi}\bra{\phi}$, then the decoherence factor (whose square
is known as the Loschmidt echo, $L(t) = \modulo{\nu(t)}^2$
\cite{losch}) is given by the overlap $\nu(t) =
\braket{\phi_1(t)}{\phi_0(t)}$, where $\ket{\phi_{\alpha}} = e^{-i
H_{\alpha} t} \ket{\phi}$.

The quantity $\nu(t)$ can be used to characterize the environment
itself, and, in particular, it gives a very peculiar behavior for
fermionic environment. Indeed, as first pointed out by P. W.
Anderson over 40 years ago, such an overlap of the two many-body
wavefunctions, describing deformed and un-deformed Fermi seas,
respectively, scales with the size of the environment and vanishes
in the thermodynamic limit, giving rise to an `orthogonality
catastrophe' \cite{anderson}. The dynamic counterpart of
Anderson's theory was investigated a few years later with the
prediction of a universal absorption-edge singularity in the X-ray
spectrum of simple metals, which has become known as the
`Fermi-edge singularity' \cite{mnd}. Mahan, Nozieres and De
Dominicis (MND) obtained an expression for the function $\nu(t)$
describing the response of a Fermi gas to the sudden switching of
a local perturbation, i.e. a core-hole induced by the X-ray, which
gives rise to a deformation (or shake up) of the many body state
of the gas.

It is our aim in this paper to study the analogous of such a
phenomenon for a trapped gas of ultra-cold Fermi atoms in which
the very fast excitation of an impurity atom (e.g. by a focused
laser pulse) produces a sudden local perturbation
\cite{catastro1,catastro2,knap}. The time response of the gas is
directly related to the decoherence of the impurity, which
experiences a purely dephasing dynamics. As mentioned above, for
such a case the non-Markovianity of the map is strictly connected
to the decoherence factor. As a result, with our theoretical
construction we are able to explore the link between the
non-Markovianity and the orthogonality catastrophe occurring
within the environment.

We will adopt the geometric measure recently put forward by some
of us in Ref. \cite{det}, which allows for an intuitive
visualization of the information exchange between system and
environment. This is briefly recalled in Sect. \ref{secuno}. Then,
after the model for the fermionic bath is explicitly described in
Sec. \ref{secdue}, in Sec. \ref{sectre} we use the geometric
measure to discuss the decoherence of the impurity. Some final
remarks are given in Sec. \ref{secquattro}.

\section{Geometric description of non-Markovianity}
\label{secuno} 
\begin{figure}[t]
\centerline{\scalebox{0.3}{\includegraphics{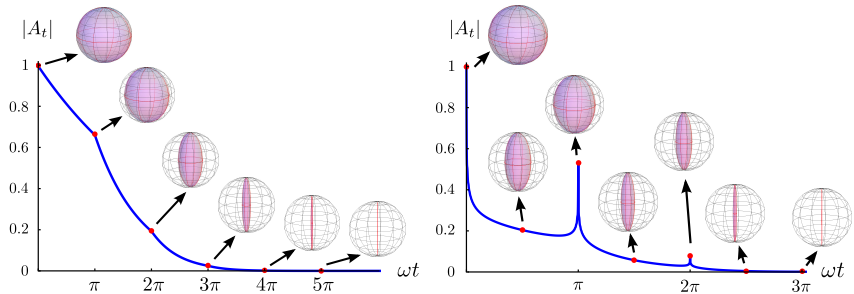}}}
\caption{Behavior of the determinant $|A_t|$, of the dynamical map
governing the time evolution of the  two-level impurity atom, for
the case of an environment with $N_{F}=200$ fermions and for
$\beta=0.05$ , $\alpha=0.001$ (left plot) and $\beta=3$ ,
$\alpha=0.1$ (right plot). (Energies are in units of $\hbar
\omega$).} \label{detpal}
\end{figure}

In this section we briefly review the definition
and meaning of the measure of non-Markovianity introduced in Ref.
\cite{det}, adapting it to the case of a qubit undergoing a purely
dephasing dynamics. The basic idea is that the time evolution of
the density matrix of a qubit can always be recast into the form
of an affine transformation for the Bloch vector $\vett b =
\mbox{Tr} \{ \vett{\sigma} \}$ (where $\vett{\sigma}$ is the
vector of Pauli matrices), which can be contracted, rotated and
translated by a given amount. In particular, for the time
evolution given in Eq. (\ref{tev}), the translation term is absent
and we have \be \vett b(t) = A_t \, \vett b(0) \, ,\ee with
\begin{equation}
A_t=\left(\begin{matrix}
\mathfrak{Re} \nu(t) & \mathfrak{Im} \nu(t) & 0 \\
- \mathfrak{Im} \nu(t) & \mathfrak{Re} \nu(t) & 0 \\
0 & 0 & 1
\end{matrix}\right) \, .
\end{equation}
The generic initial state for the qubit corresponds to a Bloch
vector lying within the unit (Bloch) sphere. The set of accessible
states changes as a function of time, being contracted (with
respect to the initial sphere) in the equatorial plane for the
case of a purely dephasing evolution. In Fig. \ref{detpal} we
provide a representation of this set as a function of time for two
specific situations which will be described in more detail in the
next section.
In the general case, the absolute value of the determinant of the
matrix describing the dynamical map, $||A_t||$, gives the ratio
between the volume of the set of states (or, more precisely, the
set of $\vett b$ vectors) accessible by the system at time $t$ and
the volume of the initial set of all possible Bloch vectors, which
is the entire Bloch sphere.

A dynamical map described by a Lindblad-like master equation gives
rise to a non-increasing volume \cite{uno,det}. This is true even
if the master equation has time decay coefficients, provided that
the latter are strictly positive quantities \cite{piilo}. Hence,
it is natural to call a process Markovian if the determinant does
not increase in time, and non-Markovian if an initial volume
contraction is followed by a temporary inflation, giving rise to a
determinant that has a positive time derivative in some specific
interval.

The two examples reported in Fig. \ref{detpal} explicitly depict a
Markovian and a non-Markovian evolution in terms of the
determinant.

Using the same method adopted in Ref. \cite{blp} to single out the
intervals in which the determinant increases in time, we define
\cite{det}
\begin{equation}
\mathcal{N}_V=\int_{\frac{d\left||A_t|\right|}{dt}>0}\frac{d\left||A_t|\right|}{dt}
\, dt \label{NMnostra}
\end{equation}
as a non-Markovianity measure.

For the map corresponding to a dephasing evolution, the
determinant is related to the decoherence factor. Explicitly,
$\left||A_t|\right|=|\nu(t)|^2$. In Section \ref{sectre}, we will
discuss the behavior of ${\cal N}_V$ for a qubit in a fermionic
environment as a function of the coupling strength and of the
temperature of the environment.

\section{Impurity in a Fermionic environment}
\label{secdue} The explicit model for the environment that we
discuss in this paper is given by a gas of ultra-cold
non-interacting fermionic atoms, trapped in an harmonic potential
of frequency $\omega$. This is described by the Hamiltonian
$$\hat{H}_{env} = \sum_{n} \en {c}_{n}^{\dagger} {c}_{n} \, ,$$
with $c_{n}$ being the annihilation operator for the $n$-th single
particle level of energy $\en = \hbar \omega (n+1/2)$.
$\hat{H}_{env}$, together with the number operator $\hat N =
\sum_{n} {c}_{n}^{\dagger} {c}_{n}$,  also sets the initial
equilibrium state of the gas, $\rho_{env} = \exp \{ - \beta ( \hat
H_{env} - \mu \hat N)\} /Z$, where the chemical potential $\mu$ is
fixed by the requirement that the gas contains (on average) $N_F$
fermions, while $\beta$ is the inverse temperature.

We consider a two-level impurity, trapped in an auxiliary
potential and brought in contact with the Fermi gas. We assume
that when the impurity is in the state $|0\rangle$, it has a
negligible scattering interaction with the gas. On the other hand,
if the impurity is in the state $\ket 1$, the gas feels a
localized perturbation $\hat V$, describing a neutral $s$-wave
like interaction that we treat in the pseudo-potential
approximation: $V(x)=\pi V_{0}x_{0}\delta (x)$ (the trap length
$x_{0}$ as well as the factor $\pi$ are put in the definition for
future convenience only). The interaction Hamiltonian, then, has
the form \be H_{int} = \hat{V} \otimes \ket{e}\bra{e} \, . \ee

As mentioned in the introduction, the key quantity for our
discussion is the decoherence factor \be \nu(t) = \left\langle
e^{\frac{i}{\hbar }\hat{H}_{env}t}{\,}e^{-\frac{i}{\hbar}
(\hat{H}_{env}+\hat{V}) t}\right\rangle \text{,} \label{nu}
\end{equation}
where the braket symbol is a short-hand notation for the thermal
equilibrium average over the unperturbed environment.

An analytic estimate can be given for this quantity both at zero
and at finite temperatures \cite{catastro2}. Here, we will proceed
to a numerical evaluation of the decoherence factor (and, in
particular, of its modulus) using the linked cluster theorem and
up to two-vertices connected Feynman diagrams, which amount to a
partial re-summation of a perturbative expansion in the ratio of
the interaction strength $V_0$ with the Fermi energy
$\varepsilon_F = \hbar \omega (N_F + \frac{1}{2})$. The details of
this kind of evaluation are given in Ref. \cite{catastro2}, where
it is shown that a good indicator of the effect of the
perturbation induced on the Fermi gas is the a-dimensional
parameter \be \alpha =\frac{V_{0}^{2}}{\hbar \omega \,
\varepsilon_F} \, .\ee This coincides with the critical parameter
of the MND theory, which is obtained as the limiting case of a
free Fermi gas, $\omega \rightarrow 0$.

In the following, we will consider spin $1/2$ fermions, fix their
number to $N_F = 200$, and discuss the behavior of the measure of
non-Markovianity ${\cal N}_V$  as a function of $\alpha$ and of
the inverse temperature $\beta = \frac{1}{KT}$.

\section{Non-Markovianity and its relation to the shake-up of the Fermi gas}
\label{sectre} 
\begin{figure}[t]
\centerline{\scalebox{0.3}{\includegraphics{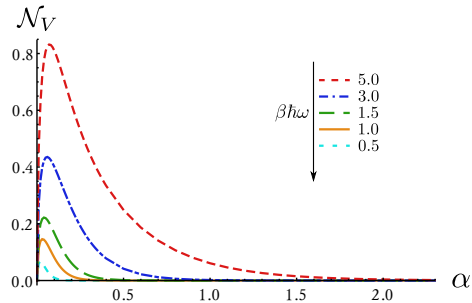}}}
\caption{Non-Markovianity measure ${\cal N}_V$ as function of the
effective coupling strength $\alpha$, for various values of the
inverse thermal energy $\beta$ expressed in units of $1/\hbar
\omega$.} \label{noncre}
\end{figure}

The parameter $\alpha$ introduced in the previous
section is a measure of the strength of the perturbation due to
the switching impurity. One could naively expect that by
increasing $\alpha$ the amount of non-Markovianity in the dynamics
of the qubit should increase. However, this is not the case as
Fig. \ref{noncre} clearly shows. In particular, ${\cal N}_V$
increases for a very small $\alpha$, reaching a maximum value for
$\alpha \leq 0.2$, which depends on the chosen value of $\beta$.
After such a maximum, ${\cal N}_V$ decreases with increasing the
interaction strength.
This is due to the peculiar way a fermionic system responds to the
perturbation, especially at small temperatures. For very small
$\alpha$, an almost linear increase with the intensity of the
perturbation is expected from simple second order perturbation
theory. Indeed, to first order in $V_0$ one would obtain only  a
shift of the energy levels, resulting in a purely oscillating
$\nu(t)$; the second order correction in $V_0$ (which are linear
in $\alpha$), instead, introduces a distortion of the
single-particle energy eigenstates.

A small $\alpha$ implies that the Fermi surface is not
substantially modified and that only those fermions whose energy
is close to the Fermi energy are excited. This, in turns, means
that only a few fermionic modes (and, thus, few almost undistorted
frequencies) enter the dynamics, and give rise to a
quasi-periodicity of the function $\nu(t)$ with frequency
$\omega$. As a result, every half a period the derivative of the
determinant changes sign and a contribution is given to the
integral in Eq. (\ref{NMnostra}). The accumulation of such
positive contributions gives rise to an ${\cal N}_V$ that grows
with $\alpha$. Notice, however, that a true periodicity is quickly
lost with increasing $\alpha$, especially if the temperature is
not kept low. From Fig. \ref{detpal}, one can see that only a
vague periodicity survives already at $\alpha=0.1$ if $KT = \hbar
\omega/3$.

This increase of ${\cal N}_V$ with increasing $\alpha$ is
counteracted by an effective suppression of the oscillations in
$\nu(t)$ which occurs when the single-particle energies become
more and more distorted (so that they are not anymore multiple of
$\omega$) and when more and more transitions are induced by the
increasing-in-strength of the perturbation. For large values of
$\alpha$, indeed, the entire Fermi sea responds to the
perturbation and the non-markovianity decreases.

Such a behavior can be also interpreted in a complementary way.
For small $\alpha$'s the effective environment felt by the
impurity has a very prominent spectral structure, given by the
Fermi edge. On the other hand, with increasing $\alpha$, the
effective environmental frequency spectrum becomes more and more
flat, giving rise to an effective Markovian dynamics for the
qubit.

This line of reasoning is confirmed by the fact that the amount of
non-Markovianity decreases with increasing the temperature due to
the fact that the Fermi edge is more and more blurred for a
smaller and smaller $\beta$.

\begin{figure}[t]
\centerline{\scalebox{0.6}{\includegraphics{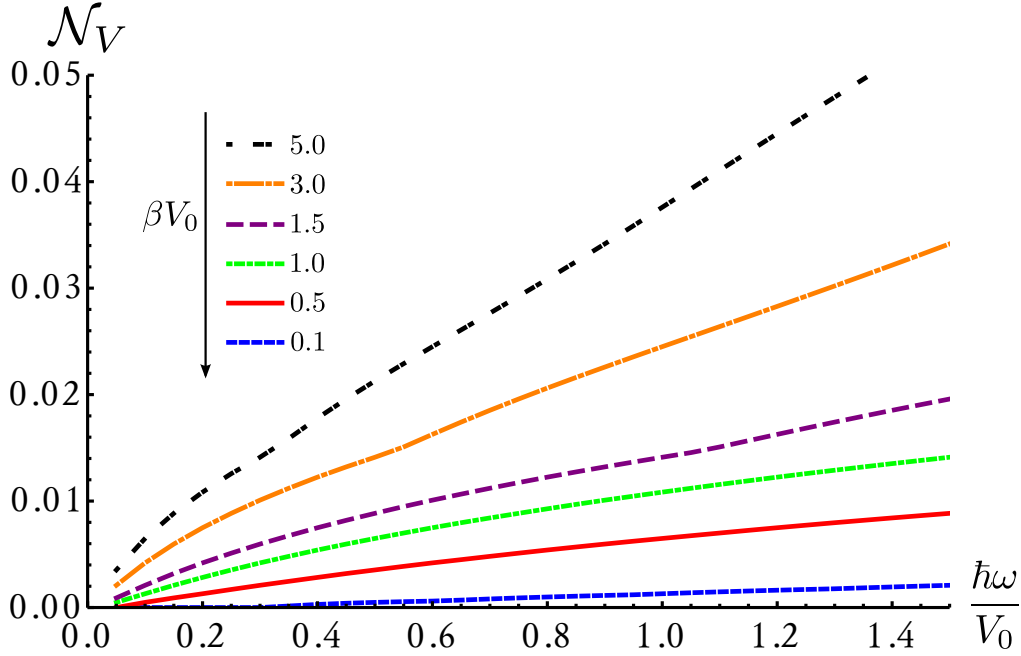}}}
\caption{Non-Markovianity measure as function of the trap
frequency $\omega$ for different temperatures.} \label{omeg}
\end{figure}

Fig. \ref{omeg} gives the behavior of the amount of
non-Markovianity as function of the trap frequency, to explicitly
confirm that ${\cal N}_V$ increases with $\omega$ as a result of
the fact that the periods in which the determinant grows become
closer to each other in time.
In particular, for the free gas originally treated by MND, which
is obtained when $\omega\rightarrow 0$, the dynamics of the
impurity is fully markovian due to the absence of any oscillations
in the decoherence factor.
\subsection{Build-up in time of the non-Markovianity}
The non-Markovianity measure ${\cal N}_V$ gives an integral
characterization of the whole dynamics. More details on the time
development of the memory effects during the system-environment
information-exchange process can be obtained if we consider
separately the time intervals in which the determinant increases
and decreases, up to a given time $t$. To this end, it is useful
to define
\begin{equation} \mathcal{N}_{\pm}(t)= \pm \int_{\pm
\frac{d\left||A_{\tau}|\right|}{d
\tau}>0}\frac{d\left||A_{\tau}|\right|}{d \tau} \, \theta(t-\tau)
\, d\tau \label{pm} \, ,
\end{equation}
which give, respectively, (the sum of ) the amount of
expansion/contraction of the volume of the set of accessible
states within the time $t$. In particular, it is meaningful to
consider the ratio \be {\cal R}(t) = \frac{N_+(t)}{ N_-(t)} \,
,\ee which expresses the fraction of the volume which is recovered
in the expansions with respect to the one lost due to the previous
contractions. The function ${\cal R} (t)$ is strictly zero until
the volume starts increasing and then it grows/diminishes with the
volume. Every time interval in which ${\cal R}$ grows corresponds
to a positive contribution to the build-up of the integral measure
${\cal N}_V$.
\begin{figure}[t]
\centerline{\scalebox{0.3}{\includegraphics{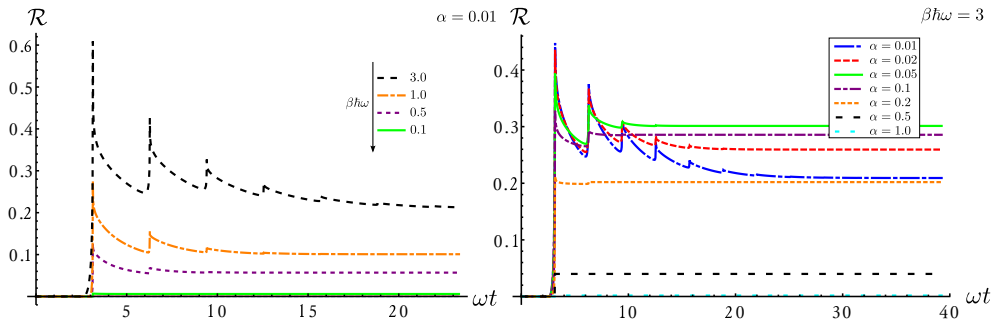}}}
\caption{Ratio expansion over contraction factors of the volume of
states accessible through the dynamical map after time $t$. The
left panel reports ${\cal R}(t)$ at fixed $\alpha=0.01$ for
various temperatures, while the right panel shows the same
quantity for different values of $\alpha$ and a fixed $KT = \hbar
\omega/3$.} \label{ratio}
\end{figure}
Fig. \ref{ratio} confirms the interpretation that the
non-Markovianity is due to quasi-periodic  oscillations of the
volume of the set of accessible states, occurring with a period $2
\pi/\omega$, which is more and more suppressed with increasing
temperature

\section{Concluding remarks}
\label{secquattro} In this paper we studied the dynamics of a
two-level impurity encapsulated in a trapped fermion environment.
Due to the presence of the Fermi edge, at low temperatures such an
environment enjoys the presence of a spectral structure which
induces a non-Markovian behavior in the time evolution of the
impurity. This is due to the oscillating response of the trapped
Fermi gas in which transition are induced by the perturbing local
impurity. Such a behavior is found to occur provided the
interaction strength is not too large (with respect to the
relevant energy scale of the gas, given by $\sqrt{\varepsilon_F
\hbar \omega}$). On the other hand, a stronger interaction tends
to smoothen the environmental spectral density giving rise to a
Markovian dynamics for the impurity. Markovianity is generically
obtained also if the temperature is larger than the scale set up
by the trapping frequency and, more generally, whenever the
discrete structure of the single-particle energy levels of the gas
can be confused with a continuum, e. g. in the absence of the
trap.

The effects that we reported can be studied with a system of
trapped cold fermionic atoms. Indeed, in this realm, many
experiments have recently dealt with the effect of impurities
within the gas~\cite{fermionexp}, which can be easily tested both
as function of the interaction strength (changeable via the
phenomenon of Feshbach resonance) and of the trapping frequency.

\end{document}